\pdfoutput=1

\documentclass[11pt,twoside,a4paper,cmspaper,final,collab]{cms-tdr}
\def\svnVersion{175299}\def\cmsCernNo{2012-372}\def\cmsCernDate{\today}\def\cmsMessage{Submitted to Physical Review Letters}

\begin{document}\cmsNoteHeader{HIG-12-041}

\hyphenation{had-ron-i-za-tion}
\hyphenation{cal-or-i-me-ter}
\hyphenation{de-vices}

\RCS$Revision: 164557 $
\RCS$HeadURL: svn+ssh://alverson@svn.cern.ch/reps/tdr2/papers/HIG-12-041/trunk/HIG-12-041.tex $
\RCS$Id: HIG-12-041.tex 164557 2013-01-14 07:58:23Z conway $

\newcommand{\Zee}{\ensuremath{\cPZ\to\Pep\Pem}}
\newcommand{\tauh}{\ensuremath{\Pgt_\mathrm{h}}}
\newcommand{\taul}{\ensuremath{\Pgt_\mathrm{\ell}}}
\newcommand{\nanob}{\mbox{{\rm ~nb}~}}
\newcommand{\fb}{\ensuremath{\mathrm{fb}}}
\newcommand{\pb}{\ensuremath{\mathrm{pb}}}
\newcommand{\ifb}{\ensuremath{\mathrm{fb^{-1}}}}
\newcommand{\ipb}{\ensuremath{\mathrm{pb^{-1}}}}
\newcommand{\Hmu}{\ensuremath{\PH\to\cPZ\cPZ^{(\ast)}\to 4\Pgm}}
\newcommand{\mmu}{\ensuremath{m_{4\Pgm}}}
\newcommand{\mH}{\ensuremath{m_{\PH}}}
\newcommand{\lnQ}{\ensuremath{\mathrm{ln}(Q)}}
\newcommand{\mlnQ}{\ensuremath{-2\,\ln(Q)}}
\newcommand{\LS}{\ensuremath{\mathcal{L}(S)}}
\newcommand{\LB}{\ensuremath{\mathcal{L}(B)}}
\newcommand{\LSB}{\ensuremath{\mathcal{L}(S+B)}}
\newcommand{\mZ}{\ensuremath{m_{\cPZ}}}
\newcommand{\Zbb}{\ensuremath{(\cPZ/\Pgg^*){\cPqb\cPaqb}}}
\newcommand{\ZZ}{\ensuremath{\cPZ/\Pgg^*\cPZ/\Pgg^*}}
\newcommand{\Zgam}{\ensuremath{\cPZ/\Pgg^*}}
\newcommand{\mumu}{\ensuremath{\Pgmp\Pgmm}}
\newcommand{\Wo}{\ensuremath{\PW}}%
\newcommand{\Wp}{\ensuremath{\PWp}}%
\newcommand{\Wm}{\ensuremath{\PWm}}%
\newcommand{\Zo}{\ensuremath{\cPZ}}%
\newcommand{\Ho}{\ensuremath{\PH}}%
\newcommand{\KD}{\mathrm{KD} }
\newcommand{\X}{\mathrm{X} }
\cmsNoteHeader{HIG-12-041}
\title{Study of the mass and spin-parity of the Higgs boson candidate via its decays to \texorpdfstring{\cPZ}{Z} boson pairs}

\date{\today}

\abstract{
A study is presented of the mass and spin-parity of the new boson
recently observed at the LHC at a mass near 125\GeV. An integrated
luminosity of 17.3\fbinv, collected by the CMS experiment in
proton-proton collisions at center-of-mass energies of 7 and 8\TeV, is
used.  The measured mass in the $\cPZ\cPZ$ channel, where both $\cPZ$
bosons decay to $\Pe$ or $\Pgm$ pairs, is $126.2 \pm 0.6 \stat\, \pm
0.2 \syst\GeV$.  The angular distributions of the lepton pairs in this
channel are sensitive to the spin-parity of the boson.  Under the
assumption of spin 0, the present data are consistent with the pure
scalar hypothesis, while disfavoring the pure pseudoscalar hypothesis.}

\hypersetup{%
pdfauthor={CMS Collaboration},%
pdftitle={Study of the mass and spin-parity of the Higgs boson candidate via its decays to Z boson pairs},%
pdfsubject={CMS},%
pdfkeywords={CMS, physics, Higgs, multilepton, diboson}}

\maketitle 

Recently the ATLAS and CMS Collaborations announced the observation of
a narrow resonance with mass near
125\GeV~\cite{Aad:2012zz,Chatrchyan:2012zz} and properties
consistent with those of the Higgs boson predicted in the standard
model
(SM)~\cite{StandardModel67_1,StandardModel67_2,StandardModel67_3} of
particle physics.  This observation may help to elucidate the nature
of spontaneous electroweak symmetry
breaking~\cite{Higgs:1964ia,Higgs:1964pj,Higgs:1966ev,Englert:1964et,Guralnik:1964eu,Kibble:1967sv}.
The main decay modes by which this resonance is observed include
photon pairs ($\Pgg\Pgg$) and massive vector boson pairs ($\PW\PW$ and
$\cPZ\cPZ$), where at least one of the vector bosons is off mass
shell.  As more proton-proton collision data are recorded at the Large
Hadron Collider (LHC), attention is turning to the determination of
various properties of this state, including its mass, spin, parity,
and couplings to SM particles.

The observation of the new boson in the $\Pgg\Pgg$ channel implies
that the resonance must be a boson with spin 0 or 2; spin 1 is
excluded by the Landau--Yang theorem~\cite{Landau,Yang}.  The decays
of the new boson to $\cPZ\cPZ$ in which both $\cPZ$ bosons decay to
charged-lepton pairs ($\ell^+\ell^-$, where $\ell = \Pe$ or $\Pgm$)
offer the possibility to probe the spin-parity and mass of the
resonance.  We describe these measurements in this Letter, using a
data set recorded by the CMS experiment in proton-proton collisions at
the LHC, corresponding to an integrated luminosity of 17.3\fbinv,
with 5.1\fbinv collected at a center-of-mass energy of 7\TeV and
12.2\fbinv at 8\TeV.

The Compact Muon Solenoid (CMS) detector, described in detail
elsewhere~\cite{Chatrchyan:2008zzk}, is a large general-purpose device
based on a silicon pixel/strip tracking system, a lead tungstate
crystal electromagnetic calorimeter (ECAL), and a brass/scintillator
hadron calorimeter, all inside the field volume of a 3.8\unit{T}
solenoidal magnet.  Outside the magnet is a multilayered muon
detection system embedded in steel absorber plates, which form the
return path for the magnetic flux, as well as forward calorimetry.
The detector is particularly well suited for measuring electron and
muon transverse momenta ($\pt$) over a wide range.

The signal candidates are selected using well-identified and isolated
prompt leptons.  The event selection and lepton reconstruction are described
elsewhere~\cite{Chatrchyan:2012zz}.  Events are selected online by
triggers requiring the presence of either an $\Pe\Pe$, $\Pe\Pgm$, or
$\Pgm\Pgm$ pair with asymmetric $\pt$ thresholds, or three electrons
with reduced thresholds.  The reconstructed electrons are required to
have $\pt^{\Pe} > 7$\GeV and to be within the tracker geometrical
acceptance, at pseudorapidities $\abs{\eta^{\Pe}}< 2.5$, where
$\eta\equiv-\ln[\tan(\theta/2)]$ in terms of the polar angle $\theta$.
The corresponding requirements for reconstructed muons are $\pt^{\Pgm}
> 5$\GeV and $\abs{\eta^{\Pgm}}< 2.4$.  The selection requires
the presence of two pairs of leptons.  The leptons in a pair must be
of opposite charge and same flavor.  Photons with $\pt^\gamma > 2$\GeV
are reconstructed within $\abs{\eta^{\gamma}} < 2.4$ and
considered as possible final-state radiation (FSR) candidates.  An FSR
photon is retained and associated with the closest lepton in a lepton
pair only if the dilepton plus photon mass is closer to the nominal
$\cPZ$ boson mass.  One lepton pair is required to be loosely
consistent with originating from a $\cPZ$ decay by demanding that the
invariant mass of the pair be in the range 40--120\GeV.  The first
pair, denoted $\cPZ_1$, is the one nearest the $\cPZ$ in mass.  The
second pair, denoted ${\cPZ_2}$, is required to satisfy $12 <
m_{\cPZ_2} < 120$\GeV.  Among the four selected leptons forming the
two $\cPZ$ boson candidates, at least one should have $\pt > 20$\GeV
and another should have $\pt > 10$\GeV.

\begin{figure}[tbh!p]
  \begin{center}
         \includegraphics[width=0.49\textwidth]{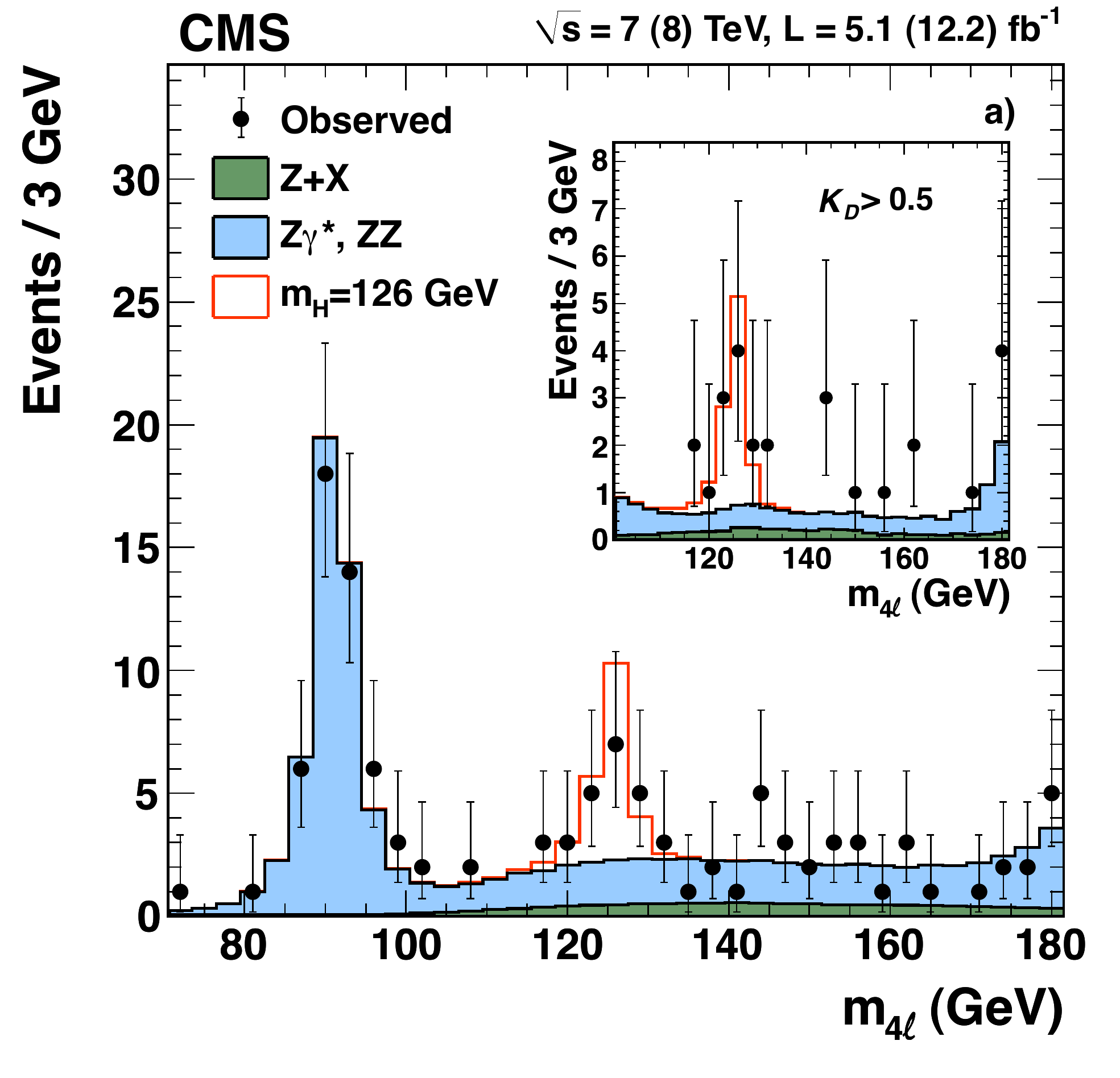}
         \includegraphics[width=0.49\textwidth]{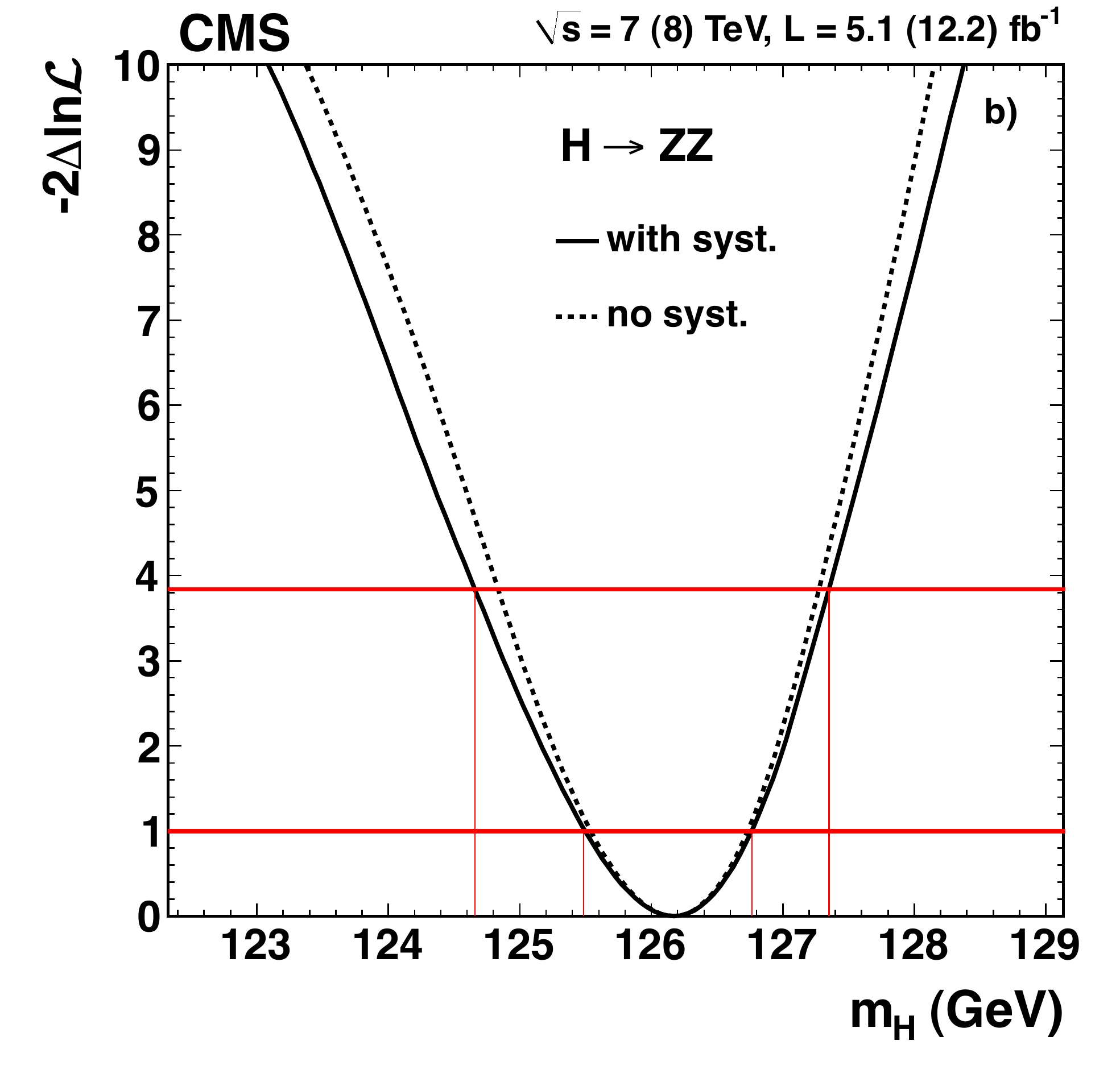}
  \end{center}
  \caption{\textit{a)} Distribution of four-lepton invariant mass in the range near the
           126 GeV resonance. Points represent the observed data, shaded histograms represent the
           backgrounds, and the open histograms represent the signal expectation.
           The inset shows the $m_{4\ell}$ distribution for events with high values of
           kinematic discriminant $K_D$.
           \textit{b)} Scan of $-2\Delta\ln \mathcal{L}$ versus $\mH$ with and without the effect
           of systematic uncertainties included. }
  \label{fig-m4l}
\end{figure}

The selected sample is dominated by continuum electroweak production
of $\cPZ\cPZ / \cPZ \gamma^*$, which constitutes irreducible background,
estimated from Monte Carlo simulation as in the previous
analysis~\cite{Chatrchyan:2012zz}.
A small background from reducible sources remains, mainly
from $\cPZ + X$ events, where $X$ consists of two reconstructed leptons,
at least one of which is a nonprompt lepton, including misidentified leptons,
leptons from heavy-quark decays, or photon conversions.
The reducible background is measured from signal-free control
regions in experimental data~\cite{Chatrchyan:2012zz}.

The performance of the signal selection and background suppression has
been improved compared with the previous
analysis~\cite{Chatrchyan:2012zz} by using a three-electron trigger,
using better muon reconstruction and momentum measurement algorithms,
fine-tuning the electron isolation requirement, and by using a
regression technique,
as previously used for the  $\PH \rightarrow \Pgg\Pgg$ analysis~\cite{Chatrchyan:2012zz},
for the contribution of the ECAL to the electron
momentum measurement.  For similar reducible background rates, the
absolute signal detection efficiency is improved by up to 4\% in the
$4\Pe$ channel and up to 2\% in the $2\Pe 2\Pgm$ channel.  The
resolution of the reconstructed mass of the $4\ell$ system is
improved, relatively, by about 10\% in the $4\Pe$ and $2\Pe 2\Pgm$
channels.  Signal candidate masses are measured with a per-event mass
precision varying between 1\% and 3\%.  The detection efficiency for a
SM Higgs boson of $\mH = 126$\GeV, with leptons within the
geometrical acceptance, is 31\% in the $4\Pe$ channel, 42\% in the
$2\Pe 2\Pgm$ channel, and 59\% in the $4\Pgm$ channel.

\begin{figure*}[thbp]
  \begin{center}
            \includegraphics[width=0.32\linewidth]{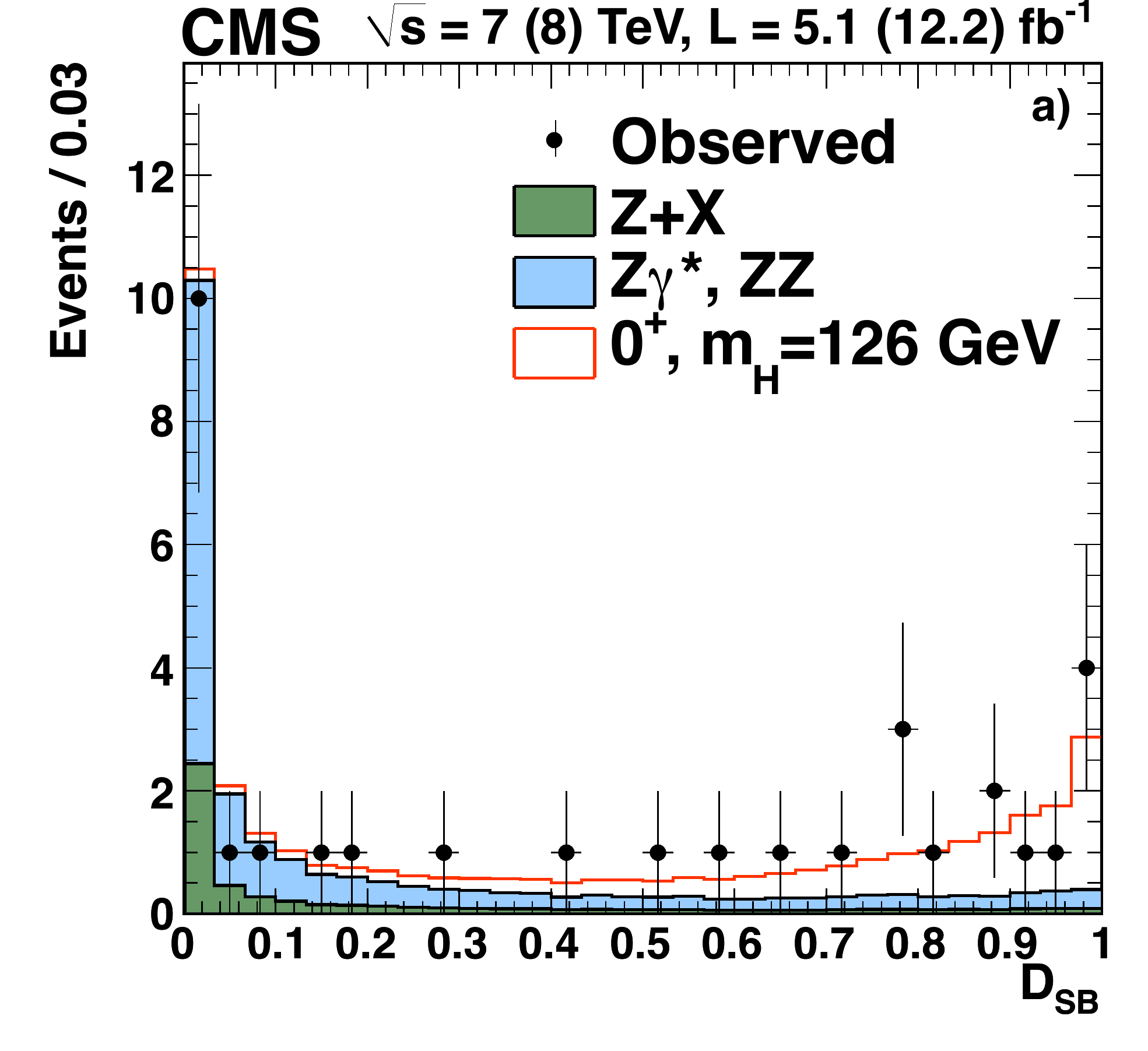}
            \includegraphics[width=0.32\linewidth]{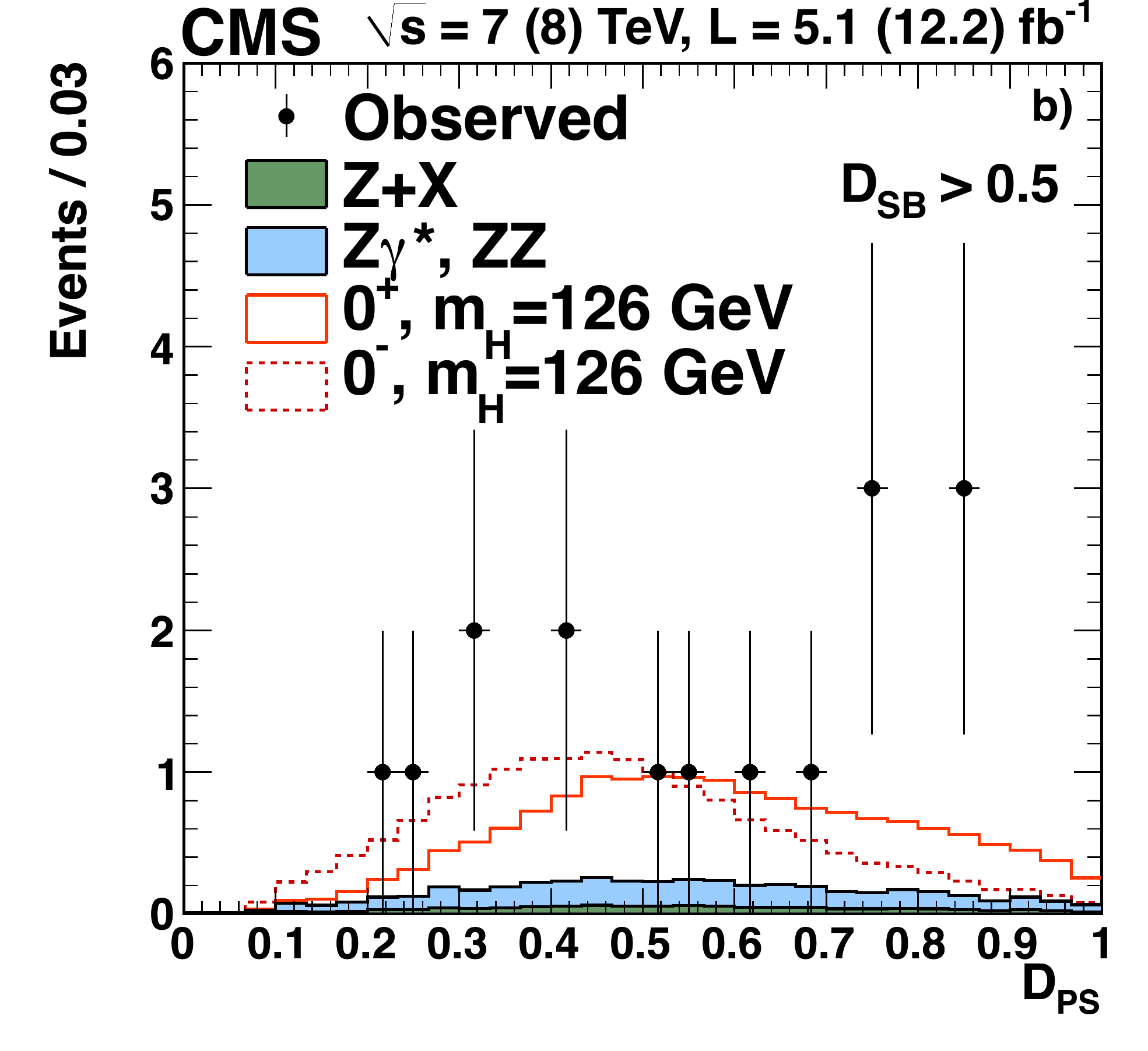}
            \includegraphics[width=0.32\linewidth]{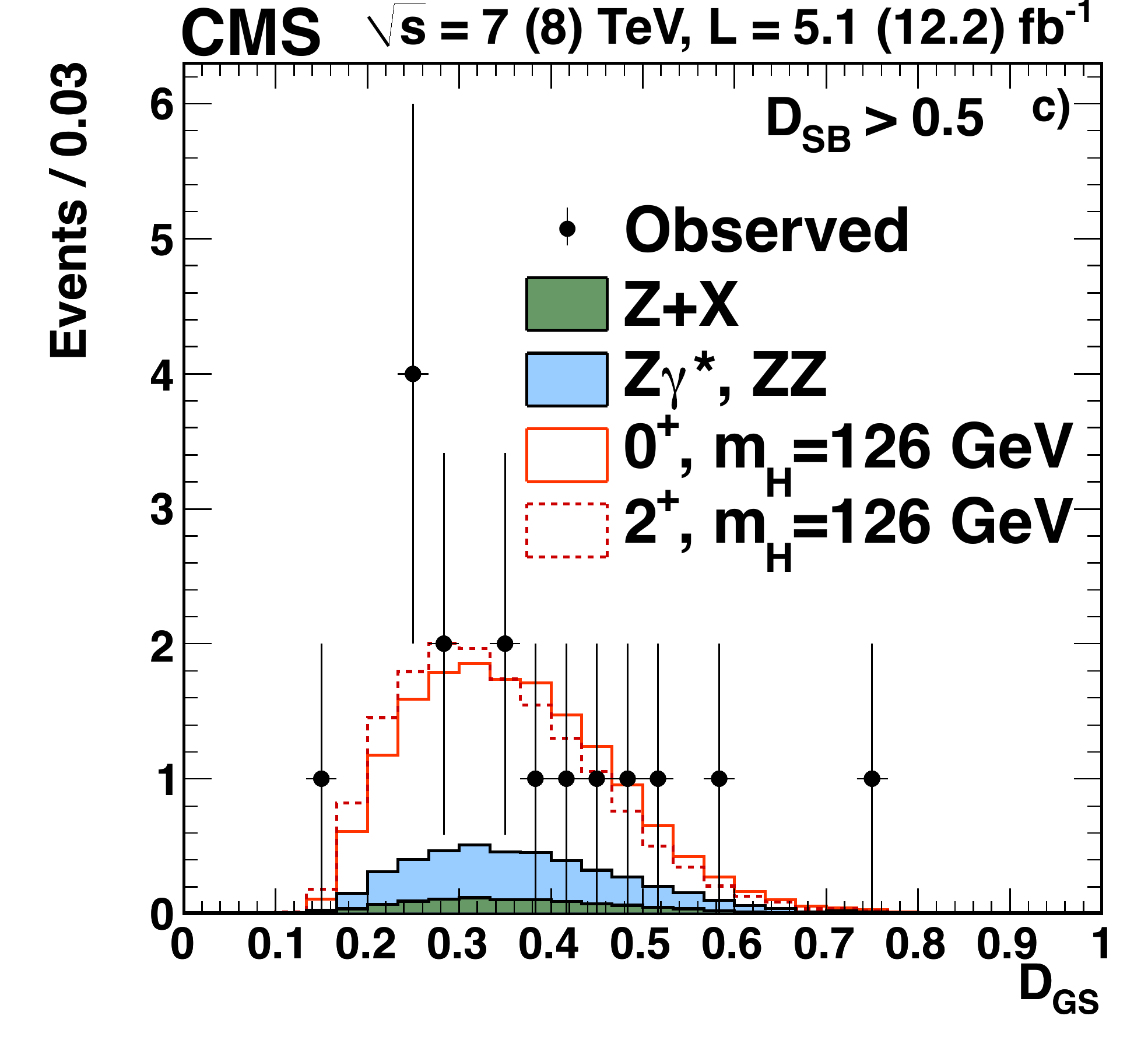}
  \end{center} \caption{
            \textit{a)}  Observed distribution of the $D_{SB}$ (SM Higgs boson versus
            background) discriminant compared with the background and
            signal expectations.
            \textit{b)} Observed distribution of
            $D_{PS}$ ($J^P=0^-$ versus $J^P=0^+$) compared with
            expectation, for $D_{SB} > 0.5$.
            \textit{c)} Observed distribution of $D_{GS}$ ($J^P=2^+$
            versus $J^P=0^+$) compared with expectation, for $D_{SB} > 0.5$.
            Points represent the observed data, shaded
            histograms represent the background, and the open
            histogram represent the expectation for a 126 GeV boson with
            the indicated spin-parity, produced at the SM Higgs boson rate.} \label{fig-MELA}
\end{figure*}

Systematic uncertainties are evaluated from the observed data for the
trigger efficiency (1.5\%) and the combined lepton reconstruction,
identification, and isolation efficiencies.  These range from 1.2\% in
the $4\Pgm$ channel to about 11\% in the $4\Pe$ channel.  Systematic
uncertainties on energy-momentum calibration and energy resolution are
incorporated through their effects on the reconstructed mass
distributions.  Uncertainties of 0.2\%, 0.2\%, and 0.1\%, are assigned
on the mass scale for the $4\Pe$, $2\Pe2\Pgm$, and $4\Pgm$ channels,
respectively.  The effect of the energy resolution uncertainties is
taken into account by incorporating a 20\% uncertainty on the
simulated width of the signal mass peak.  To validate the level of
accuracy with which the absolute mass scale and resolution are 
known~\cite{Chatrchyan:2012zz,Chatrchyan:2012-ZZ4l},
we use $\cPZ \to \ell\ell$, $\Upsilon \to \ell\ell$, and
$\JPsi \to \ell\ell$ events. The limited statistical precision of the
control samples is included as a systematic uncertainty on the final
results.  Since the reducible background is derived from control
regions, its prediction is independent of the uncertainties on the
integrated luminosity.  The integrated luminosity uncertainty (2.2\%
at 7\unit{TeV}~\cite{CMS-PAS-SMP-12-008} and 4.4\% at
8\unit{TeV}~\cite{CMS-PAS-LUM-12-001}) enters the evaluation of the
expected $\cPZ\cPZ$ background and signal rates.  Systematic
uncertainties on the Higgs boson cross section (about 18\%) and
branching fraction (2\%) are taken from
Refs.~\cite{LHCHiggsCrossSectionWorkingGroup:2011ti,Denner:2011mq}.

Figure~\ref{fig-m4l}a shows the invariant mass distribution of the
selected four-lepton events in the mass range $70 < m_{4\ell} <
180$\GeV.  The contribution expected from a SM Higgs boson of mass
$\mH =$ 126\GeV is displayed.  The peak from $\cPZ \to
4\ell$ decay, studied in detail elsewhere~\cite{CMS:2012bw}, is
observed at the nominal $\cPZ$ boson mass.  The signal from the new
boson is a distinct peak above the expected background, consistent
with the signal lineshape depicted in the figure.  The background is
locally flat and dominated by the $\cPZ\cPZ / \cPZ \gamma^*$
contribution.
In the mass range,  $121.5 < m_{4\ell} < 130.5$\GeV, corresponding to the three
central bins around the new boson peak in Fig.~\ref{fig-m4l}a, we observe 
17 events: there are 6, 8 and 3 events in the $4\Pgm$, $2\Pe2\Pgm$, and $4\Pe$ 
final states respectively. 
This compares to an expectation of $ 6.8 \pm  0.8 \stat\, \pm 0.3 \syst$ 
from SM background.

Further separation between the signal and background is provided by a
discriminant $K_D$ that incorporates the production and decay kinematics.
In this analysis, we make use of observables defined for each event in the
$4\ell$ center-of-mass frame; the rapidity and transverse momentum of the
$4\ell$ system depend on the production mechanism and are ignored.
We use a matrix element likelihood
approach~\cite{Chatrchyan:2012zz, Gao:2010qx, Bolognesi:2012, VegaMorales2012},
which combines, for each value of $m_{4\ell}$, the
two dilepton masses $m_{\cPZ_1}$ and $m_{\cPZ_2}$ and five angular
variables denoted $\vec{\Omega}$.  We introduce a kinematic
discriminant $K_D$ using the probability density in the dilepton
masses and angular variables, ${\cal
P}(m_{\cPZ_1},m_{\cPZ_2},\vec{\Omega}|m_{4\ell})$.  The discriminant
is defined as
\begin{equation}
  K_D \equiv \frac{ \mathcal{P}_\text{sig}}{ \mathcal{P}_\text{sig} + \mathcal{P}_\text{bkg}}
                          = \left[ 1 + \frac{ \mathcal{P}_\text{ bkg}(m_{\cPZ_1},m_{\cPZ_2},\vec{\Omega}|m_{4\ell}) }
                                  { \mathcal{P}_\text{sig}(m_{\cPZ_1},m_{\cPZ_2},\vec{\Omega}|m_{4\ell}) } \right]^{-1} \ \ \  .
\end{equation}
A scalar SM Higgs boson is assumed for the signal.
The separation between the signal and background is relatively insensitive to the particular
choice of  a signal spin-parity hypothesis~\cite{Bolognesi:2012}.
The minimum $p$-value~\cite{Beringer:2012zz}, which characterizes the probability for a
background fluctuation to be at least as large as the observed maximum
excess around $\mH \simeq 126\GeV$, is obtained from the measurements
of $m_{4\ell}$ and $K_D$.
It corresponds to a significance of 4.5 standard deviations, which is
to be compared to an expected significance of 5.0 standard deviations
for the SM Higgs boson.

We measure the mass of the boson using a maximum-likelihood fit
to three-dimensional distributions combining for each event the
$m_{4\ell}$, the associated per-event uncertainties $\delta
m_{4\ell}$~\cite{Chatrchyan:2012-ZZ4l} calculated from the individual
lepton momentum errors, and $K_D$.
The signal strength $ \Pgm$ (defined below) is a free parameter in
this mass fit.
A scalar SM Higgs boson is assumed for the signal lineshape.
Figure~\ref{fig-m4l}b shows the
value of $-2\Delta\ln\mathcal{L}$, where $\mathcal{L}$ is the likelihood, as
a function of $\mH$, with and without the effects of systematic
uncertainties included.  An estimate for the mass of $126.2 \pm
0.6 \stat\, \pm 0.2 \syst\GeV$ is obtained.  Combined with the result
from the $\Pgg\Pgg$ channel~\cite{Chatrchyan:2012zz}, we obtain a mass
of $125.8 \pm 0.4 \stat\, \pm 0.4 \syst\GeV$.  This value improves
upon and supersedes the previous result.

We then compare the observations with the expectation for the SM Higgs
boson at the mass value fixed to 125.8\GeV, and obtain a
measurement of the signal strength $ \Pgm= \sigma/\sigma_\mathrm{SM}$,
the production cross section times the branching fraction relative to
the SM expectation.  This is evaluated from a scan of a profile
likelihood ratio.  We perform an unbinned maximum-likelihood fit of
the two-dimensional distributions $ \mathcal{P}(m_{4\ell}|\mH) \times
\mathcal{P}(K_D | m_{4\ell})$ for the signal, and $ \mathcal{
P}(m_{4\ell}) \times \mathcal{P}(K_D | m_{4\ell})$ for the background.
The fit is performed simultaneously in the $4\Pe$, $2\Pe 2\Pgm$, and
$4\Pgm$ channels.  We obtain a signal strength of $\Pgm = 0.80^{+
0.35}_{-0.28}$, consistent with the expectation for a SM Higgs boson.

The kinematics of the production and decay of the new boson in the
$\cPZ\cPZ \to 4\ell$ channel are sensitive to its spin and
parity~\cite{Soni:1993jc, Barger:1993wt, Choi:2002jk, Allanach:2002gn,
Buszello:2002uu, Godbole:2007cn, Keung:2008ve, Antipin:2008hj,
Hagiwara:2009wt, Gao:2010qx, DeRujula:2010ys, Gainer:2011xz,
Bolognesi:2012, VegaMorales2012}.
To distinguish any two spin-parity hypotheses, we use discriminants of
the form $\mathcal{D}_{12} = \mathcal{P}_1 / ( \mathcal{P}_1 + \mathcal{P}_2)$,
where $\mathcal{P}_1$ and $\mathcal{P}_2$ are the probability densities in
$m_{\cPZ_1}$ , $m_{\cPZ_2}$, and $\vec{\Omega}$ corresponding to the
two spin parity hypotheses we wish to discriminate and include
parametrizations of the $m_{4\ell}$ distribution for a resonance at
the mass of the new boson.  We define two spin-parity discriminants:
$D_{PS}$ for the discrimination between a SM Higgs boson and a pure
pseudoscalar state $J^P=0^-$; $D_{GS}$ for discrimination between a SM
Higgs boson and a spin-two tensor state $J^P=2^+$ with the minimal
graviton-like coupling to gluons in production and to $\cPZ$ bosons in
decay.
We also define a discriminant
$D_{SB} = \mathcal{P}_\text{sig} / ( \mathcal{P}_\text{sig} + \mathcal{P}_\text{bkg})$,
similar to $K_D$ but where the probability densities also include
$m_{4\ell}$,  for the discrimination between a SM Higgs boson,
with $J^P=0^+$, and the background.

We then fit the observed data in a two-dimensional plane of $D_{PS}$
or $D_{GS}$ versus $D_{SB}$ in the mass range $106 < m_{4\ell} < 141$
GeV and obtain the likelihood values $\mathcal{L}_1$ and $\mathcal{L}_2$ for
two hypotheses of each signal type plus background.
Figure~\ref{fig-MELA}a shows the observed projections of $D_{SB}$ for
events in this mass range, and for a SM Higgs boson signal with
$\mH = 126$\GeV.  Figures~\ref{fig-MELA}b and \ref{fig-MELA}c
show the projections of the $D_{PS}$ and $D_{GS}$ discriminants, for
events with $D_{SB} > 0.5$.  In these latter two cases, the
distributions for the spin-parity states being distinguished are also
illustrated in the plot.  More data are needed for significant
discrimination of the $0^+$ from the $2^+$ hypothesis.

\begin{figure}[thbp]
  \begin{center}
     \includegraphics[width=0.49\textwidth]{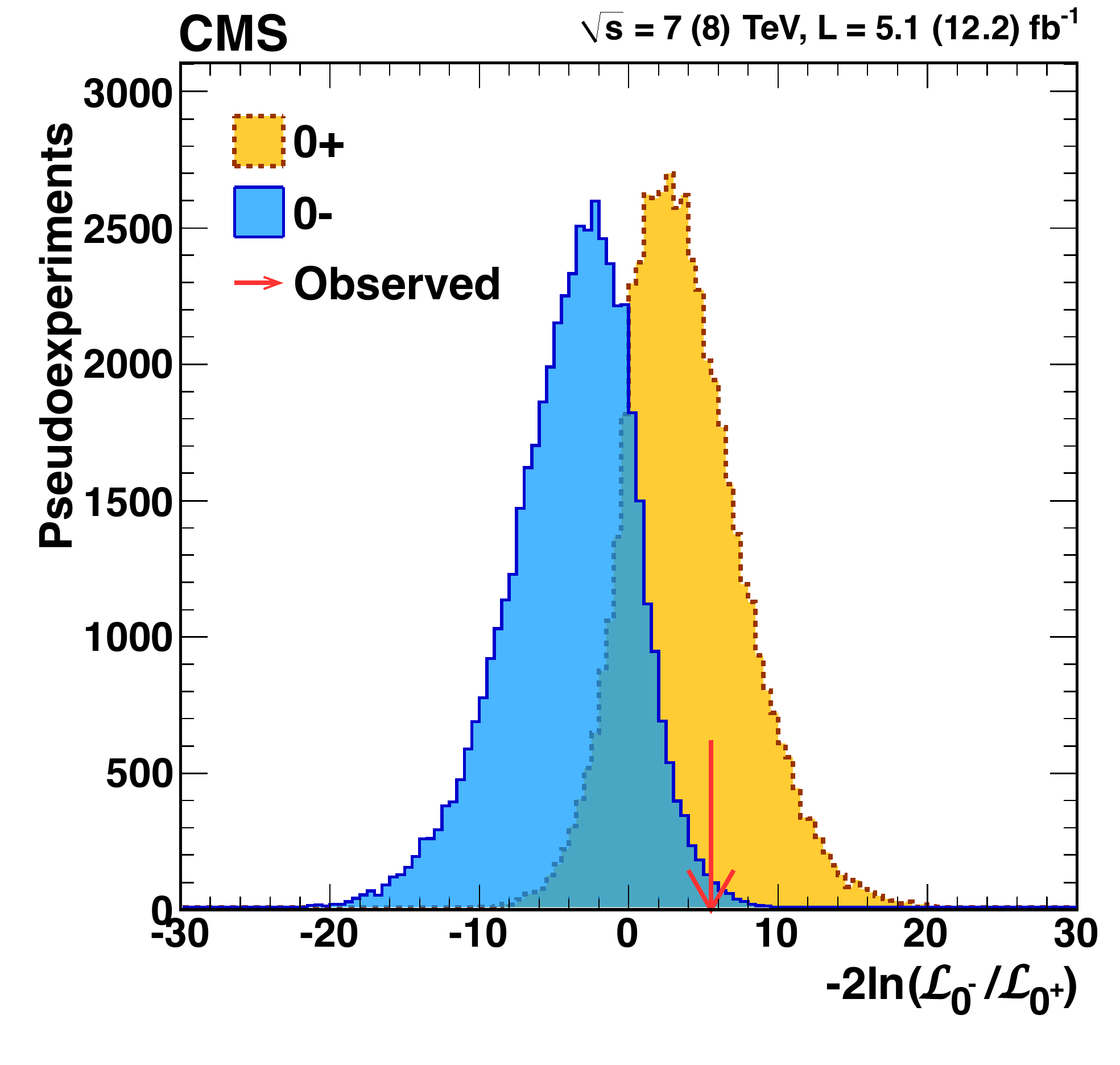}
  \end{center}
  \caption{Expected distribution of $-2\ln\mathcal{L}_{0^-}/\mathcal{L}_{0^+}$ under
           the pure pseudoscalar and pure scalar hypotheses (histograms).
           The arrow indicates the value determined from the observed data.  }
  \label{fig-LR}
\end{figure}

Figure~\ref{fig-LR} shows the distributions of the log-likelihood
ratio $-2\ln \mathcal{L}_{0^-}/\mathcal{L}_{0^+}$ from pseudoexperiments
under the assumptions of either a pure scalar or a pure pseudoscalar
model.  The arrow indicates the observed value.
Under the assumption of spin 0, the test statistic formed from a profile likelihood ratio
$\lambda = \mathcal{L}_{0^-}/\mathcal{L}_{0^+}$ of the $0^-$ and $0^+$ hypotheses
yields a $p$-value of 0.072\% for $0^-$ and a $p$-value of 0.7 for $0^+$,
with $-2\ln\lambda = 5.5$ favoring $0^+$.
This corresponds to a CL$_{\rm s}$~\cite{Read-2002} value of 2.4\%,
a more conservative value for judging whether the observed data are 
compatible with $0^-$.
The results presented here
have been confirmed with independent methods~\cite{Avery:2012um} based
on leading-order matrix elements~\cite{Alwall:2007st}.

In summary, we have measured the mass of the new boson to be
$126.2 \pm 0.6 \stat\, \pm 0.2 \syst\GeV$ in the $\cPZ\cPZ$ channel,
where both $\cPZ$ bosons decay to lepton pairs.
Combining results from the $\Pgg\Pgg$ and $\cPZ\cPZ$ channels, we obtain a 
mass of $125.8 \pm 0.4 \stat\, \pm 0.4 \syst\GeV$, which improves upon previously 
published results. 
At this mass the signal strength $ \Pgm= \sigma/\sigma_\mathrm{SM}$
is measured to be $\Pgm = 0.80^{+0.35}_{-0.28}$.
Under the assumption of spin
zero, the observed data are consistent with the pure scalar
hypothesis, while disfavoring the pure pseudoscalar hypothesis.  This
is the first study of the spin-parity of the newly discovered boson.

We congratulate our colleagues in the CERN accelerator departments for
the excellent performance of the LHC and thank the technical and
administrative staffs at CERN and at other CMS institutes for their
contributions to the success of the CMS effort. In addition, we
gratefully acknowledge the computing centers and personnel of the
Worldwide LHC Computing Grid for delivering so effectively the
computing infrastructure essential to our analyses. Finally, we
acknowledge the enduring support for the construction and operation of
the LHC and the CMS detector provided by the following funding
agencies: BMWF and FWF (Austria); FNRS and FWO (Belgium); CNPq, CAPES,
FAPERJ, and FAPESP (Brazil); MEYS (Bulgaria); CERN; CAS, MoST, and
NSFC (China); COLCIENCIAS (Colombia); MSES (Croatia); RPF (Cyprus);
MoER, SF0690030s09 and ERDF (Estonia); Academy of Finland, MEC, and
HIP (Finland); CEA and CNRS/IN2P3 (France); BMBF, DFG, and HGF
(Germany); GSRT (Greece); OTKA and NKTH (Hungary); DAE and DST
(India); IPM (Iran); SFI (Ireland); INFN (Italy); NRF and WCU (Korea);
LAS (Lithuania); CINVESTAV, CONACYT, SEP, and UASLP-FAI (Mexico); MSI
(New Zealand); PAEC (Pakistan); MSHE and NSC (Poland); FCT (Portugal);
JINR (Armenia, Belarus, Georgia, Ukraine, Uzbekistan); MON, RosAtom,
RAS and RFBR (Russia); MSTD (Serbia); SEIDI and CPAN (Spain); Swiss
Funding Agencies (Switzerland); NSC (Taipei); ThEP, IPST and NECTEC
(Thailand); TUBITAK and TAEK (Turkey); NASU (Ukraine); STFC (United
Kingdom); DOE and NSF (USA).

\bibliography{auto_generated}   

\cleardoublepage \appendix\section{The CMS Collaboration \label{app:collab}}\begin{sloppypar}\hyphenpenalty=5000\widowpenalty=500\clubpenalty=5000\textbf{Yerevan Physics Institute,  Yerevan,  Armenia}\\*[0pt]
S.~Chatrchyan, V.~Khachatryan, A.M.~Sirunyan, A.~Tumasyan
\vskip\cmsinstskip
\textbf{Institut f\"{u}r Hochenergiephysik der OeAW,  Wien,  Austria}\\*[0pt]
W.~Adam, E.~Aguilo, T.~Bergauer, M.~Dragicevic, J.~Er\"{o}, C.~Fabjan\cmsAuthorMark{1}, M.~Friedl, R.~Fr\"{u}hwirth\cmsAuthorMark{1}, V.M.~Ghete, N.~H\"{o}rmann, J.~Hrubec, M.~Jeitler\cmsAuthorMark{1}, W.~Kiesenhofer, V.~Kn\"{u}nz, M.~Krammer\cmsAuthorMark{1}, I.~Kr\"{a}tschmer, D.~Liko, I.~Mikulec, M.~Pernicka$^{\textrm{\dag}}$, D.~Rabady\cmsAuthorMark{2}, B.~Rahbaran, C.~Rohringer, H.~Rohringer, R.~Sch\"{o}fbeck, J.~Strauss, A.~Taurok, W.~Waltenberger, C.-E.~Wulz\cmsAuthorMark{1}
% ...
\vskip\cmsinstskip
\dag:~Deceased\\
1:~~Also at Vienna University of Technology, Vienna, Austria\\
2:~~Also at CERN, European Organization for Nuclear Research, Geneva, Switzerland\\
% ...

\end{sloppypar}
\end{document}